\documentclass[12pt,a4paper]{article}
\usepackage{graphicx}
\usepackage{times}
\title{\bf  Multiple stellar generations in massive star forming complexes}
\author{J. S. Clark$^1$, B. Davies$^2$ and  M. A. Thompson$^3$  
\\
\vspace{1cm}\\
\normalsize $^1$ Dept. Physics \& Astronomy, Open University, Milton Keynes, UK\\ 
\normalsize $^2$ School of Physics \& Astronomy, University of Leeds, Leeds, UK\\
\normalsize $^3$ Centre for Astrophysics Research, University of Hertfordshire, Hatfield, UK}
\date{\mbox{}}
\begin{document}
\maketitle
\pagestyle{empty}
%
%
\def\bull{\vrule height .9ex width .8ex depth -.1ex}
\makeatletter
\def\ps@plain{\let\@mkboth\gobbletwo
\def\@oddhead{}\def\@oddfoot{\hfil\tiny\bull\quad
``The multi-wavelength view of hot, massive stars''; 39$^{\rm th}$ Li\`ege Int.\ Astroph.\ Coll., 12-16 July 2010 \quad\bull}%
\def\@evenhead{}\let\@evenfoot\@oddfoot}
\makeatother
%
%
\def\beginrefer{\section*{References}%
\begin{quotation}\mbox{}\par}
\def\refer#1\par{{\setlength{\parindent}{-\leftmargin}\indent#1\par}}
\def\endrefer{\end{quotation}}
%
%
{\noindent\small{\bf Abstract:} The formation of massive stars is an outstanding problem in stellar evolution.  However, it is expected 
that they are (predominantly) born in heirarchical environments within  massive young clusters,
 which in turn are located within larger star 
forming complexes that reflect the underlying structure of the natal molecular cloud. 
Initial observations of such regions suggest that multiple generations of stars and proto-stars are 
present, necessitating a multiwavelength approach to yield a full (proto-)stellar census; in this contribution 
we provide an overview of just such an observational approach for  Galactic examples, focusing on the G305 complex. 
}
%
%
\section{Introduction}
Imaging of external galaxies reveals that stellar formation  yields large star cluster complexes of 10s-100s of parsec in 
size, and $>>10^4$M$_{\odot}$ in integrated mass. These are luminous across the electromagnetic spectrum;  with emission at 
radio wavelengths from ionised gas, far-IR \& submm from cold molecular material, IR from heated dust, optical-UV from the  
stellar population and X-rays from both pre-MS  and massive stars. Therefore a multiwavelength approach is required 
to understand the ecology of 
such regions - and hence infer masses for unresolved regions from their integrated spectral energy distributions (SEDs) - as well as  the evolution
 of massive ($>$40M$_{\odot}$) stars from cold molecular cores through to the  Main Sequence. The latter goal is
 particularly important, since our knowledge of this process suffers from few current observational contraints and 
yet very massive stars play an inordinate role in the excitation of their environment via their UV radiation field and   wind energy.
Consequently, in order to address these interelated issues we are undertaking such a  study of  several Galactic star forming regions, 
of which the G305 complex  is of particular interest given current estimates for its
stellar content (Clark \& Porter 2004). In this contribution we briefly review the observational
dataset acquired for it as a result of this program  and highlight some initial results arising from it.

\section{The G305 star forming complex}
Located in the Scutum Crux arm at a distance of $\sim$4~kpc the G305 star forming complex appears to be one of the most massive such
 regions in the Galaxy, with the radio luminosity alone suggesting the presence of $>$30 canonical O7 V stars (Clark \& Porter 2004). 
Morphologically, it appears as a tri-lobed wind blown bubble  with a maximal extent of $\sim$30~pc
 centred on the Young Massive Clusters Danks 1 \&2 (Fig. 1). Vigorous ongoing  star formation   is present on the periphery 
of the region as evidenced by significant IR-radio emission and the presence of numerous masers (Sect. 2.2). 

\subsection{The recent star formation history of G305}
The location of ongoing star formation within the complex is indicative of triggered, sequential  activity initiated by Danks 1 \& 2. 
The presence of at least one Wolf-Rayet - the WC star WR48a - suggests that star formation must have been underway for at least
 $\sim$2.5~Myr while, following the arguments presented in Clark et al. (2009) for W51,  the lack of a population of RSGs suggests 
an upper limit to the duration of the `starburst' of $\leq$10~Myr. In order to more fully constrain the properties of Danks 1 \& 2  
and hence to determine whether they could have triggered the subsequent generations of  star formation, we have
undertaken near-IR imaging \& spectroscopic observations  of them with the HST \& VLT/ISAAC and  present a subset of the data focusing on
 Danks 1 in Figs. 2 \& 3.  

A full analysis of these data will be provided in Davies et al. (in prep.) but we highlight that both clusters appear to have 
integrated masses $>>$10$^3$M$_{\odot}$. Surprisingly, given their apparent proximity (a {\em projected} separation of $\sim$3.5~pc) 
there appears to be a notable age difference ($\sim$2-3~Myr) between them, evident in both the spectral types of cluster members and the 
location of the Main Sequence  turn on. Focusing on Danks 1, we identify a number of emission line objects with spectra consistent 
with O Iafpe/WN7-9h stars; the cluster being reminiscent of the Arches in the Galactic Centre. The presence of such stars is of 
interest since they  are expected to be  massive core-H burning objects in which  very high mass loss rates cause them to 
present a more evolved  spectral type. Combined with their prodigious UV-fluxes, they are likely to be significant sources of feedback 
and  detailed non-LTE model atmosphere analyses of these objects is currently underway in order to quantify this. 
In contrast such stars are absent in Danks 2, with the presence of a WC star and O supergiants of a later spectral type 
indicating an older spectral population. 

However, massive (post-)MS objects are not restricted to these clusters. As well as the dusty WC star WR48a,
 recent IR observations have located a further 3 WC 
and 1 WN stars within the wind blown bubble (Shara et al. 2009, Mauerhan, van Dyk \& Morris 2009), suggesting that an additional dispersed
population is present within the complex, although their origin - e.g.  ejected from a cluster or formed {\em in 
situ} - is uncertain.  In this regard it closely resembles 
30 Dor, which Walborn \& Blades (1997) showed  hosts a young central cluster and   a diffuse, older population distributed across 
the wind blown cavity with   an additional (pre-MS) component located on the periphery.
Massive stars also appear present on the perimeter of G305, with
Leistra et al. (2005) showing that  the young cluster found within the cavity G305.254+0.204 contains at least
one  early O star. Moreover, early OB pre-MS stars are also found in the bubble PMN1308-6215 to the NW of the complex; the 
spectrum presented in Fig. 4 being dominated by H\,{\sc i} line  and CO bandhead emission, indicative of a hot ionising source 
surrounded by a cool accretion disc/torus, respectively. A full presentation and analysis of these and other data on the pre-MS 
population of G305 will be provided in Clark et al. (in prep.). 

\subsection{Earlier phases of (triggered) star formation}

We next turn to the more deeply embedded massive protostars and the reservoir of cold molecular material. The former may be 
identified with ultracompact H\,{\sc ii} regions,  very bright mid-far IR sources and H$_2$O \&  methanol maser 
emission, while the latter may be mapped via molecular tracers such as NH$_3$ or sub-mm continuum emission from cold ($\leq$50K) dust. 
Hill et al. (2006)
presented a survey of cold dust for selected regions within G305, finding a total of $\sim$23,000M$_{\odot}$ 
material 
located in clumps with masses up to $\sim$4,500M$_{\odot}$, although it is  expected that these will comprise lower mass 
subclumps at higher spatial resolution. Recently, Hindson et al. (2010) undertook a molecular survey of the whole complex which 
revealed a total reservoir  of cold gas of $\sim$6$\times10^5$M$_{\odot}$ (Fig. 5); even allowing for a relatively low star formation efficiency
($<10$\%) this is sufficient to yield a substantial stellar population. In order to provide a higher resolution map of this material and to determine
its properties such as clump mass function and temperature, 
we have obtained both APEX/LABOCA and Herschel far-IR - submm observations. A preliminary reduction of the 870$\mu$m   LABOCA data 
is provided in Fig. 6, which shows the `skeleton' of cold molecular 
 material upon which current 
star formation is occuring (see Clark et al. in prep. \& Thompson et al. in prep. for a full analysis). 

Finally, SEDs constructed from the full near-IR to sub-mm datasets allows the identification of Massive Young Stellar Objects (MYSOs) via 
their characteristic colours (e.g. Hoare et al. 2005), as well as a determination of their integrated bolometric luminosities. 
We show the location of such MYSOs in a subfield of G305
in  Fig. 7 (as well as H$_2$O and methanol masers; Hindson et al. 2010). Clearly significant star 
formation that will result in a new population of massive  stars is currently underway, and appears to be located on the surface of the 
molecular cloud adjacent to  the  nearby stellar cluster, suggesting that it has been 
triggered by the action of the OB stars contained within.

\section{Concluding remarks}
 
A multiwavelength approach to the study of star forming complexes allows us to locate the different stellar populations within these regions
 and hence determine the propagation (or otherwise) of star formation through the host GMC. Model atmosphere analysis 
of the  massive  stellar population constrains the feedback from such stars as well as helping date the onset of star formation via comparison to theoretical evolutionary 
tracks. Full analysis of 
the near-far IR  SEDs of embedded sources yields their bolometric luminosity and hence an  estimate of  mass, while the 
far-IR - sub-mm SED  will play a similar role for cold molecular cores - the first stage of massive star 
formation. Finally a synthesis of these data will provide a complete census of star formation within the cluster complex, an 
estimate of the efficiency of this process and - via comparison of the mass functions of the differing populations - constraints on the physics governing GMC fragmentation and subsequent cluster/star formation. 

\begin{figure}[h]
\centering
\includegraphics[angle=270,width=12cm]{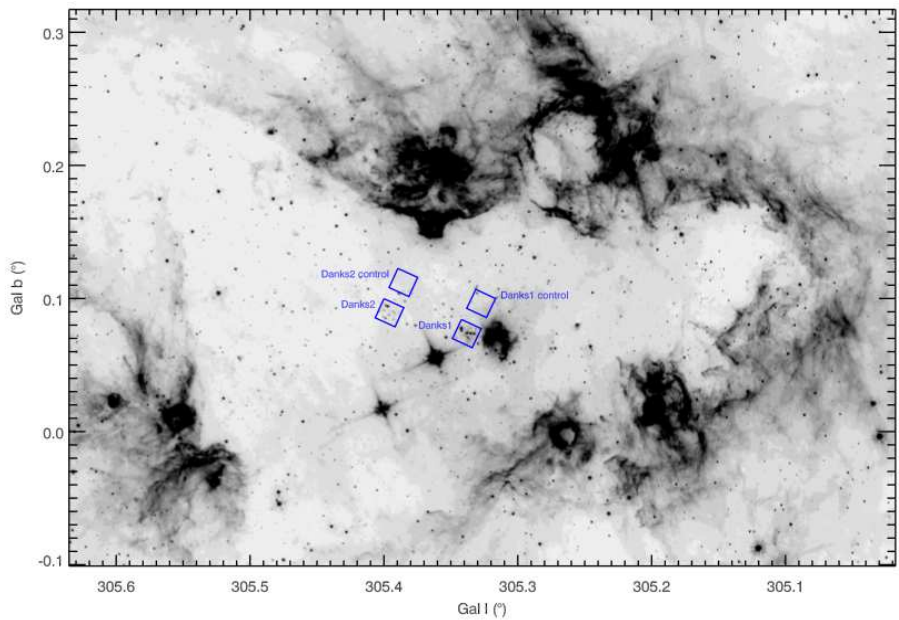}
\caption{5.8$\mu$m image of the G305 complex with the HST target  fields
for Danks 1 \& 2 indicated.}
\end{figure}

\begin{figure}[h]
\begin{minipage}{8cm}
\centering
\includegraphics[width=8cm]{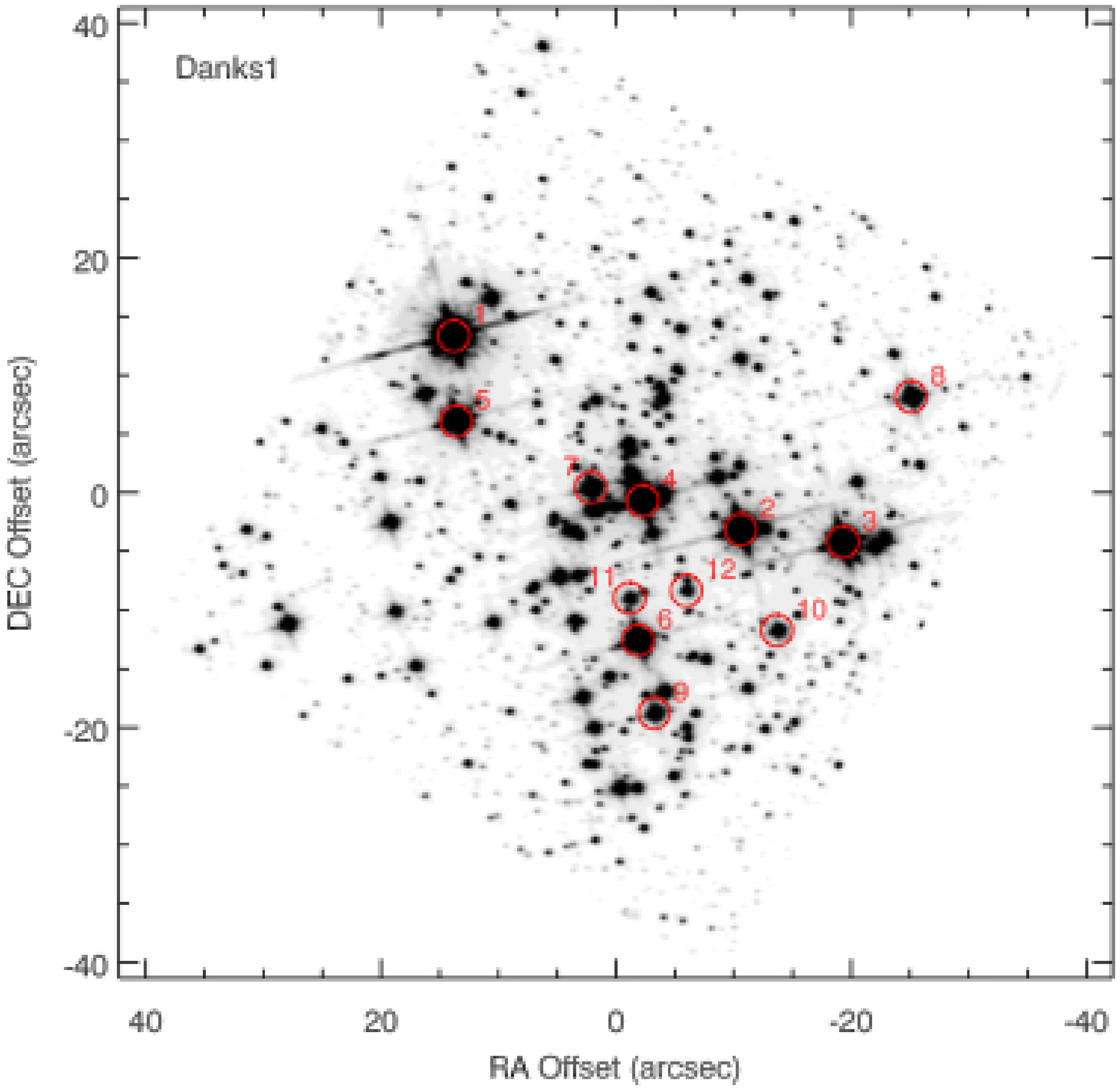}
\caption{F160W mosaic of Danks 1}
\end{minipage}
\hfill
\begin{minipage}{8cm}
\centering
\includegraphics[width=8cm]{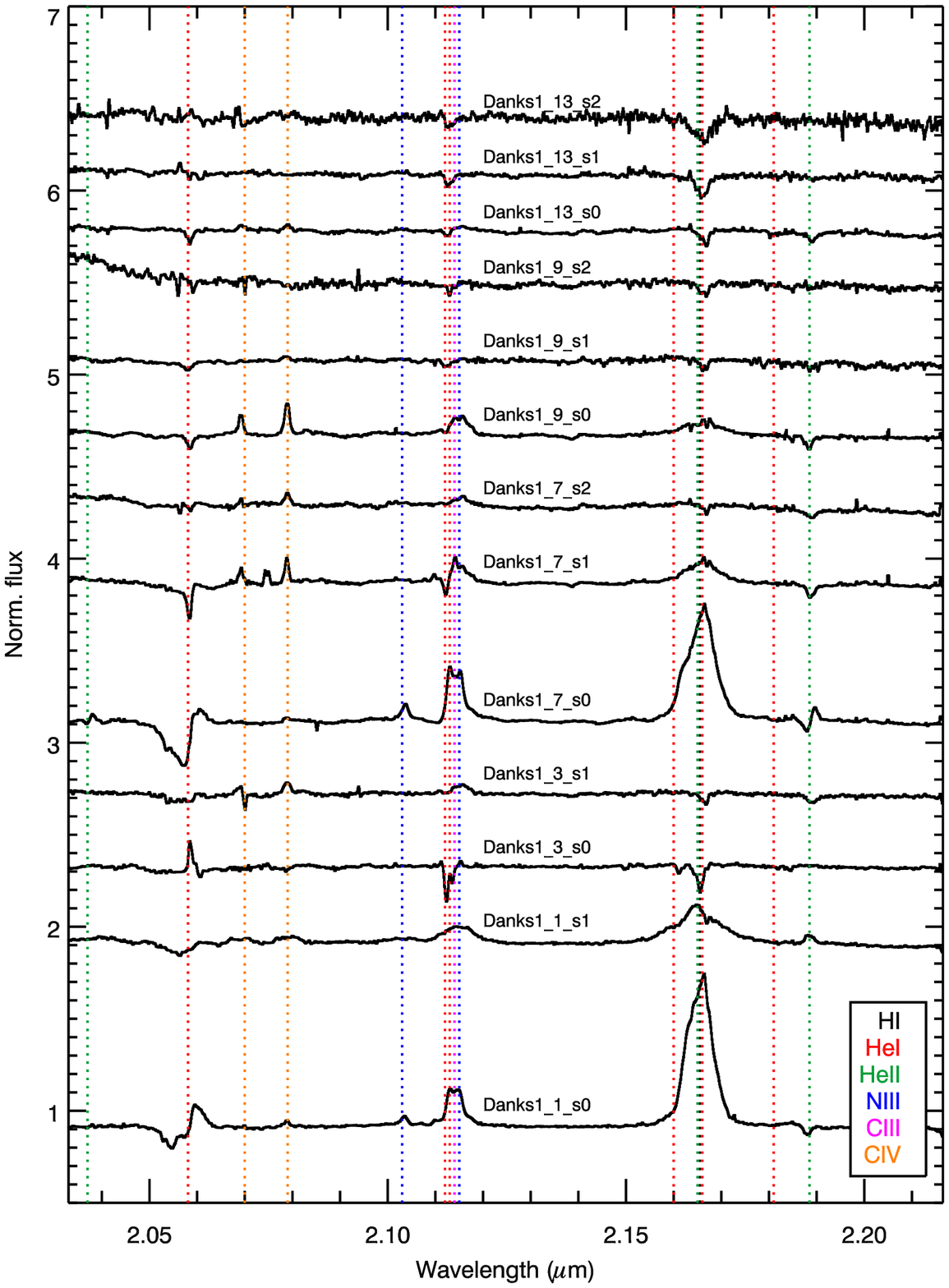}
\caption{K band spectra of massive stars within Danks 1 (Davies et al. in prep.).}
\end{minipage}
\end{figure}

\begin{figure}[h]
\begin{minipage}{6cm}
\centering
\includegraphics[angle=0,width=6cm]{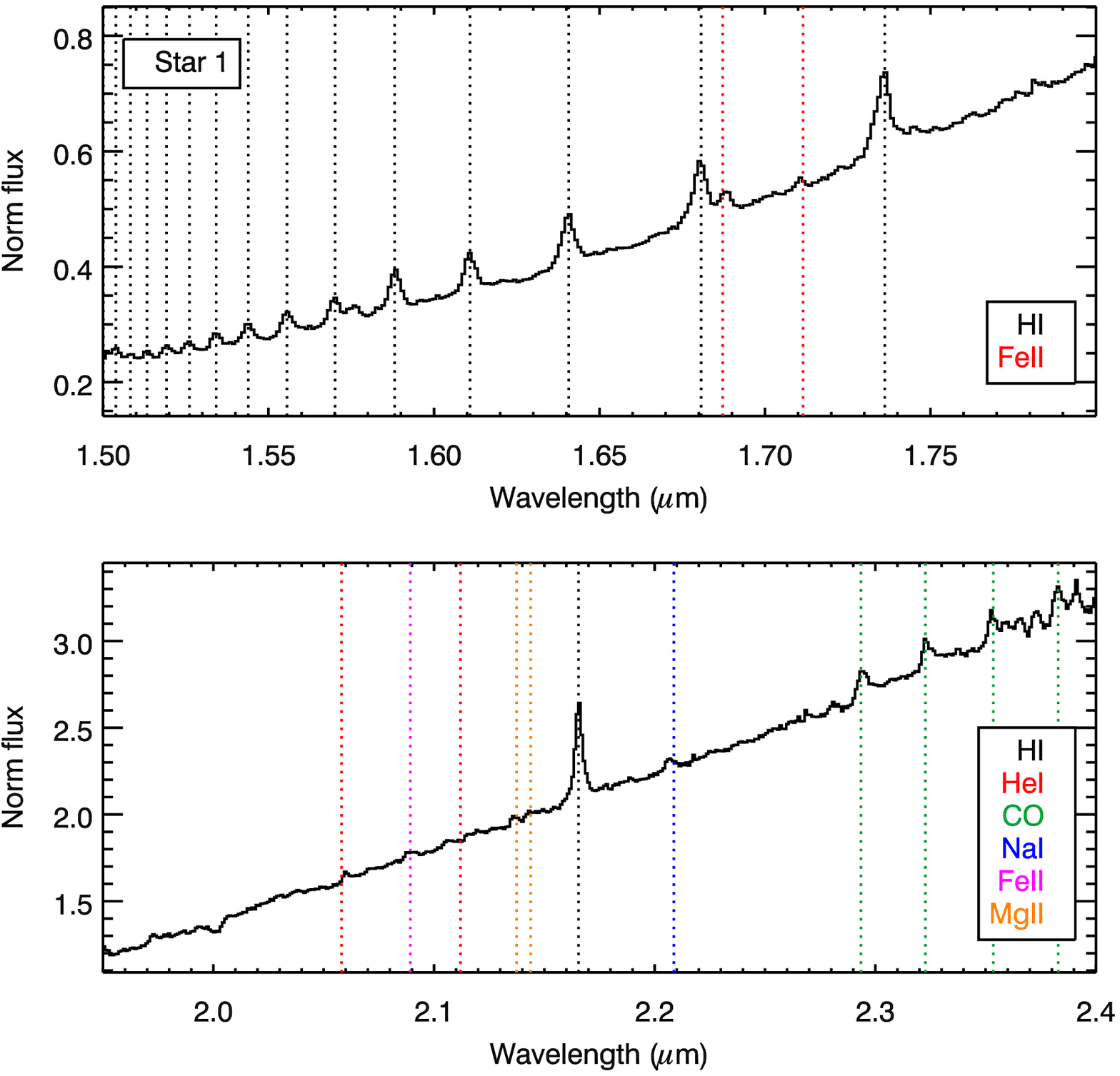}
\caption{H and K band spectra of a pre-MS OB star in the star forming bubble PMN 1308-6215
to the NW of the complex proper.}
\end{minipage}
\hfill
\begin{minipage}{10cm}
\centering
\includegraphics[angle=270,width=10cm]{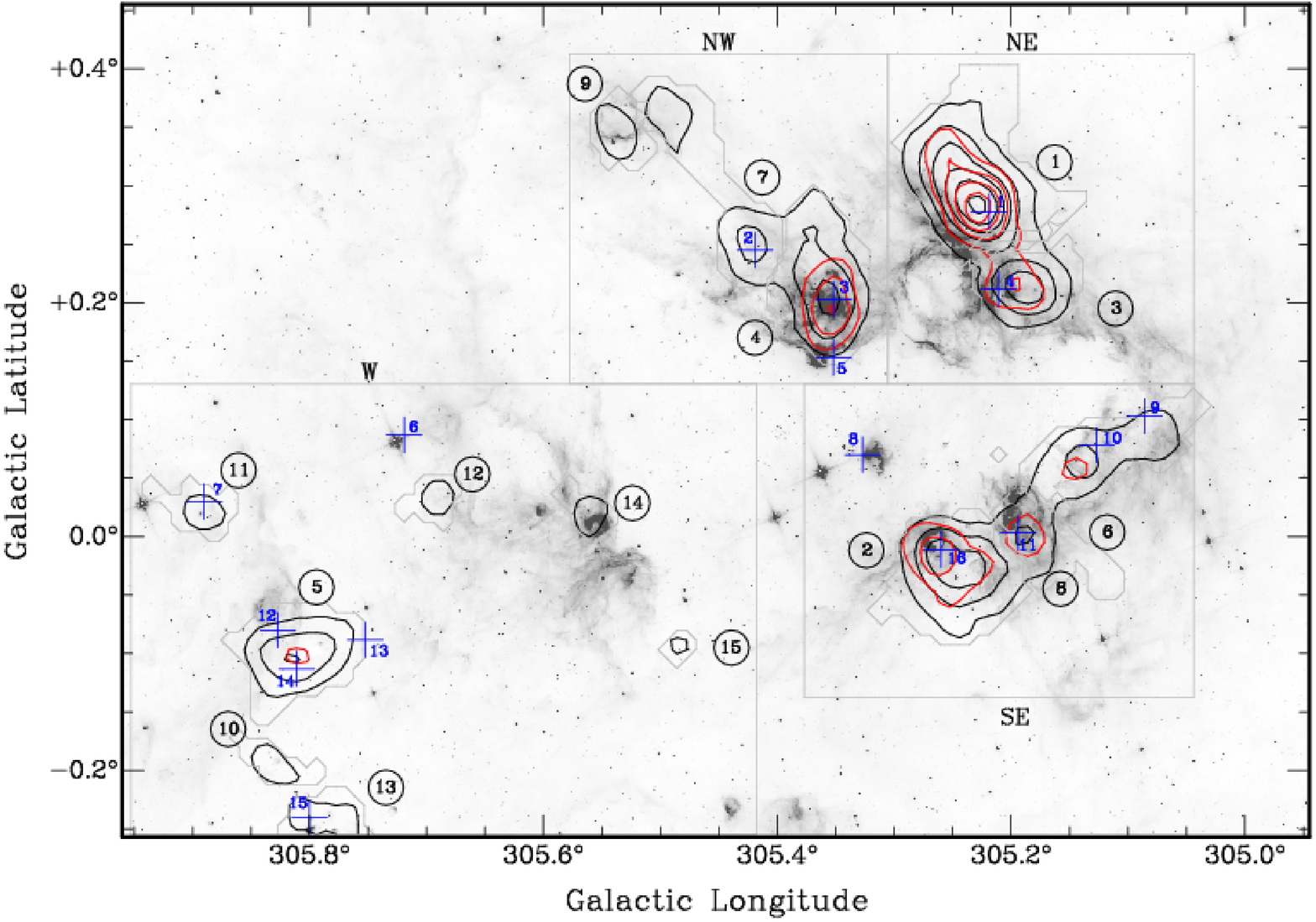}
\caption{Overplot of NH$_3$ contours on a 5.4$\mu$m greyscale image. H$_2$O masers indicated by 
crosses (Hindson et al. 2010).}
\end{minipage}
\end{figure}

\begin{figure}[h]
\begin{minipage}{10cm}
\centering
\includegraphics[angle=270,width=10cm]{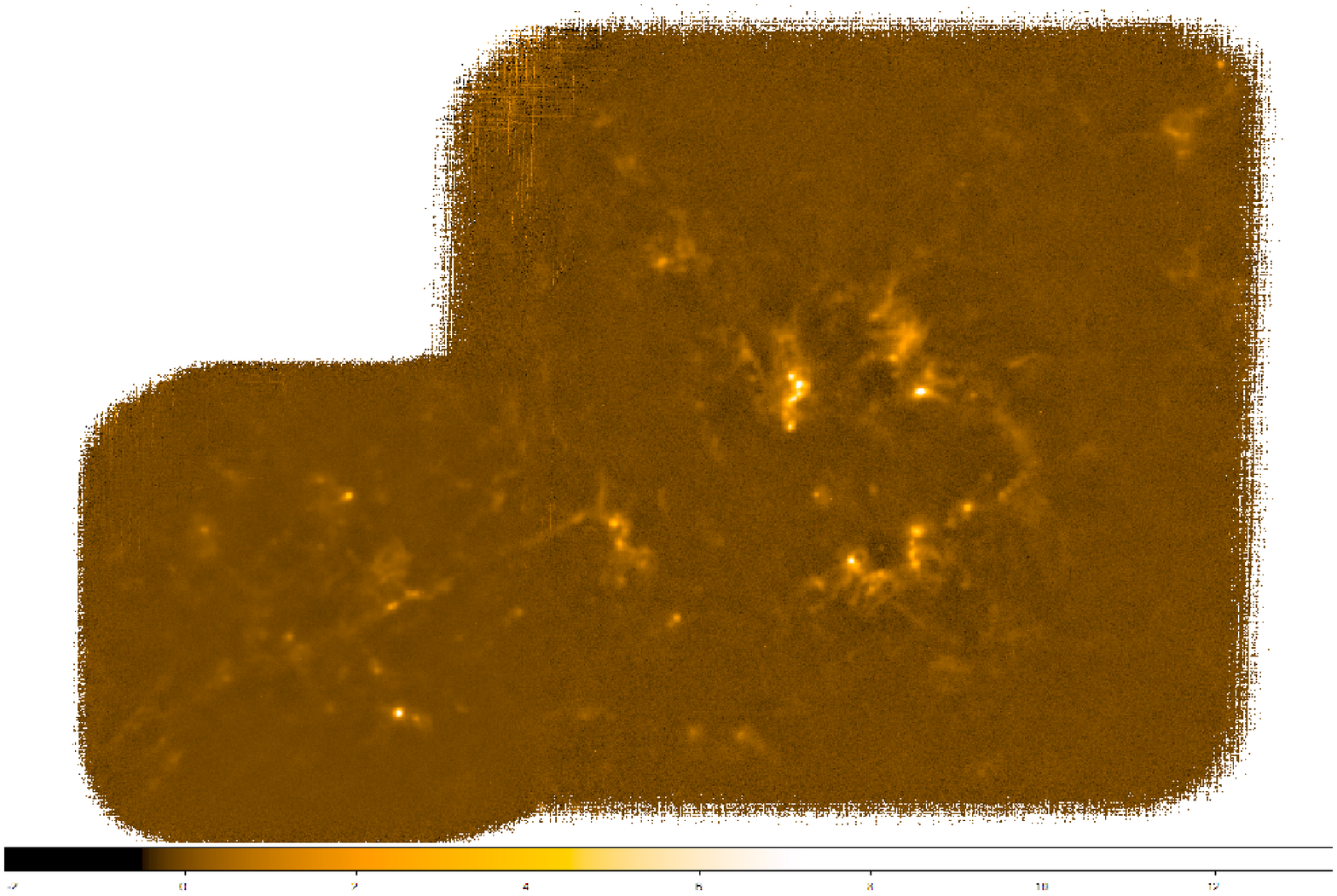}
\caption{Preliminary reduction of APEX/Laboca 870$\mu$m  continuum observations of the G305 complex.}
\end{minipage}
\hfill
\begin{minipage}{6cm}
\centering
\includegraphics[width=6cm]{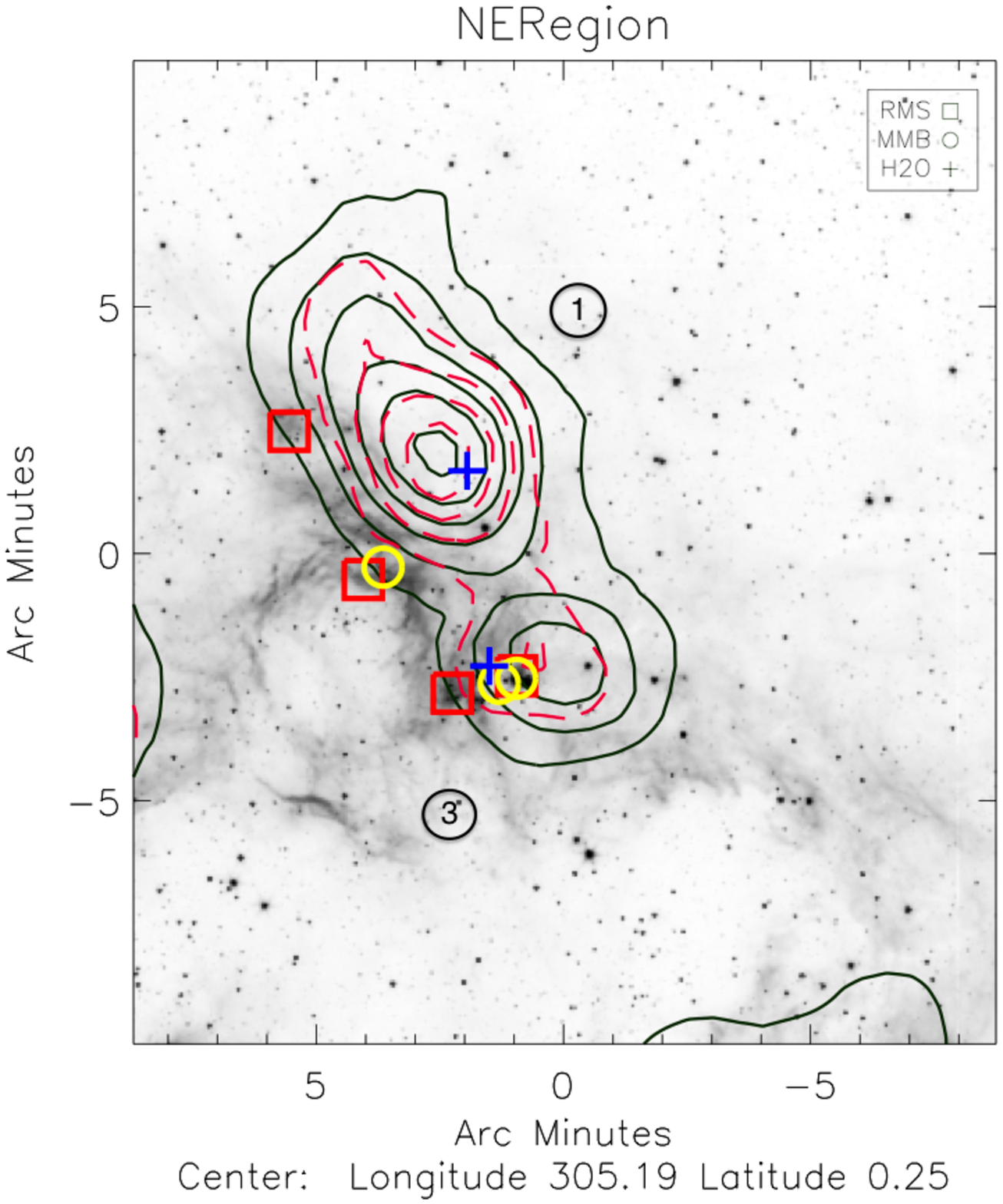}
\caption{Blow up of the NE region of the complex from Fig. 5. Crosses and contours have the same 
meaning while squares represent MYSOs and circles methanol masers (Hindson et al. 2010).}
\end{minipage}
\end{figure}

%
%
\footnotesize
\beginrefer

\refer Clark J., Porter, J., 2004, A\&A, 427, 839

\refer Clark J., Davies B., Najarro F., et al., 2009, A\&A, 504, 429

\refer Hill T., Thompson, M., Burton M., et al. 2006, MNRAS, 368, 1223

\refer Hindson L., Thompson M., Urquhart, J., Clark, J., Davies B., 2010, MNRAS, 408, 1438

\refer Hoare M., Lumsden S., Oudmaijer R., et al., 2005, IAUS 227, 370

\refer Leistra, A., Cotera A., Liebert J., Burton M., 2005, AJ, 130, 1719

\refer Mauerhan J., van Dyk S., Morris P., 2009, PASP, 121, 591 

\refer Shara M., Moffat A., Gerke J., et al., 2009, AJ, 138, 402

\refer Walborn N., Blades C., 1997, ApJS, 112, 457

\endrefer           
\end{document}